	\documentclass[aps,prb,twocolumn, showpacs,preprintnumbers,
		amsmath, amssymb, floatfix,10pt,groupedaddress]{revtex4-2} 
	\usepackage{bm}
	\usepackage{color}
	\usepackage{float}
	\usepackage{dcolumn}
	\usepackage{graphicx}
	\usepackage{booktabs}
	\usepackage{multirow}
	\usepackage{longtable}

	\newcolumntype{d}[1]{D{.}{.}{#1}}

	\usepackage[dvipsnames]{xcolor}

\begin{document}


\title{Binding energies of excitonic complexes in type-II quantum
  rings from diffusion quantum Monte Carlo calculations}

\author{D.\ M.\ Thomas}
\email{d.thomas4@lancaster.ac.uk}

\author{R.\ J.\ Hunt}

\author{N.\ D.\ Drummond}

\author{M.\ Hayne}

\affiliation{Department of Physics, Lancaster University, Lancaster
  LA1 4YB, United Kingdom}

\date{\today}


\begin{abstract}
Excitonic complexes in type-II quantum-ring heterostructures may be considered
as artificial atoms due to the confinement of only one charge-carrier type in
an artificial nucleus. Binding energies of excitons, trions, and biexcitons in
these nanostructures are then effectively ionization energies of these
artificial atoms. The binding energies reported here are calculated within the
effective-mass approximation using the diffusion quantum Monte Carlo method and
realistic geometries for gallium antimonide rings in gallium arsenide. The
electrons form a halo outside the ring, with very little charge density inside
the central cavity of the ring. The de-excitonization and binding energies of
the complexes are relatively independent of the precise shape of the ring.
\end{abstract}


\maketitle



\section{Introduction}

Quantum-dot and quantum-ring heterostructures have long been hailed as
``artificial atoms'' 
\cite{Ashoori_1996a,Gammon_2000,Warburton_2000,Bayer_2000} due to
their ability to confine charge carriers in all three spatial
dimensions. Material combinations exhibiting type-I band alignment
produce nanostructures in which both electrons and holes are confined
to the same spatial region, and such nanostructures have been studied
extensively over the last two decades 
\cite{Banin_1999,Fuhrer_2001,Nagasawa_2013}. In type-II
nanostructures, on the other hand, only holes but not electrons (or
\textit{vice versa}) are confined, presenting a rich variety of new
physics \cite{Bansal_2008,Bansal_2009,Kroutvar_2004a}. For example,
GaSb quantum dots or quantum rings in GaAs provide very deep confining
potentials for holes \cite{Nowozin_2012}, while strain in the GaSb
raises the conduction-band minimum, expelling the electrons
\cite{Hayne_2003}. Excitonic complexes in type-II nanostructures are
in fact very much more like artificial atoms than is the case for
type-I nanostructures, because the electrons are bound to the holes in
the ``artificial nuclei'' purely by the Coulomb interaction, rather
than being confined themselves. Type-II quantum rings are an
intriguingly distinct type of artificial atom with no natural
analog due to the radical difference between the ring-shaped ``artificial
nucleus'' and the pointlike nucleus of a real atom.

Excitons in type-II quantum dots have
been extensively studied both
experimentally \cite{Suzuki_1999,Hatami_1998,Geller_2003,Laghumavarapu2007} and
theoretically \cite{Williamson_2000,Andreev_2000,Nowozin_2012}; however, while
there has been some experimental work on carrier complexes in type-II
quantum-ring
nanostructures \cite{Carrington_2013,Young_2012,Young_2014,Hodgson_2016}, there
has been little theoretical work to date. The spatial separation of charge
carriers allows for a variety of interesting optoelectronic
properties \cite{Carrington_2013,Timm_2010}, including extended recombination
times, making type-II quantum rings ideal candidates for applications such as
memory devices \cite{Hayne_2013} and solar cells \cite{Laghumavarapu2007}.
Binding energies of excitonic complexes reported here are effectively
ionization energies of these artificial atoms. GaSb
quantum rings in GaAs may be produced by molecular beam
epitaxy \cite{Young_2012,Timm_2010,Hodgson_2016,Kobayashi_2004} and can form
with a variety of different cross-sections ranging from triangular, to
semicircular \cite{Young_2012}, and even trapezoidal \cite{Timm_2010}. These
quantum rings exhibit type-II behavior, with the holes strongly confined to the
rings.  Scanning tunneling microscopy (STM) has been used to investigate the
shape and size of the GaSb rings, and their optical properties have been
studied in photoluminescence
experiments \cite{Carrington_2013,Young_2012,Young_2014,Hodgson_2016}.

In this work we solve an effective-mass model of excitons (X),
positive and negative trions (X$^+$ and X$^-$), and biexcitons (XX) in
type-II quantum-ring heterostructures, focusing on GaSb rings in
GaAs. The holes are confined to the ring, which is modeled as an
infinite potential well, while the electrons are excluded from the
ring but bound to the holes by an isotropically screened Coulomb
interaction. The kinetic energy of the tightly confined holes is the
dominant contribution to the total energy of each complex; however,
the electron-hole attraction is non-negligible, as is the hole-hole
repulsion. The ring was chosen to have a rectangular cross-section for
computational convenience. The ring is centered on the origin,
orientated so that the axis of rotation is the $z$-axis and the
midpoint in the $z$ direction is the $x$-$y$ plane. The three
parameters defining the ring geometry are the half height of the ring
$R_z$, the inner radius of the ring $r_{\rm i}$, and the outer radius
$r_{\rm o}$. In our model the electron and hole densities do not
overlap, so we cannot estimate recombination rates; however our model
is reasonable for calculating binding energies.

Energies are given in units of the exciton Rydberg $R_{\rm
  y}^* = \mu e^4/[2(4\pi\epsilon)^2\hbar^2]$ and lengths in units of
the exciton Bohr radius $a_0^* = 4\pi\epsilon\hbar^2/(\mu e^2)$, where
$\epsilon$ is the permittivity of the medium, $\hbar$ is the Dirac
constant, and $e$ is the magnitude of the electron charge. The
electron-hole reduced mass is $\mu=m_{\rm e}^*m_{\rm h}^*/(m_{\rm
  e}^*+m_{\rm h}^*)$, where $m_{\rm e}^*$ and $m_{\rm h}^*$ are the
effective masses of an electron and a hole, respectively.  Within the
effective-mass approximation the Hamiltonian for a biexciton is
\begin{eqnarray}
\frac{\hat{H}}{2R_{\rm y}^*} &= &-\frac{(a_0^*)^2\mu}{2m_{\rm
    e}^*}\left(\nabla_{\rm e_1}^2+\nabla_{\rm e_2}^2\right)-
\frac{(a_0^*)^2\mu}{2m_{\rm h}^*}\left(\nabla_{\rm h_1}^2+\nabla_{\rm
  h_2}^2\right) \nonumber \\& & {} +\frac{a_0^*}{r_{\rm
    e_1e_2}}+\frac{a_0^*}{r_{\rm h_1h_2}}-\frac{a_0^*}{r_{\rm e_1h_1}}-
\frac{a_0^*}{r_{\rm e_1h_2}}-\frac{a_0^*}{r_{\rm e_2h_1}}-\frac{a_0^*}{r_{\rm
    e_2h_2}} \nonumber
\\& & {} +\sum_i{V_i\left(\frac{\mathbf{r}_i}{a_0^*}\right)}
\end{eqnarray}
in excitonic units, where $r_{ij}=|\mathbf{r}_i-\mathbf{r}_j|$ and
$V_i$ is the confining potential, which is infinite inside the ring
and zero outside for electrons, and \textit{vice versa} for
holes. This is an inhomogeneous four-body problem, but the diffusion
quantum Monte Carlo (DMC) method \cite{Ceperley_1980,Foulkes_2001} can
be used to calculate the exact ground-state energy for each complex,
and hence the de-excitonization and binding energies.  The trion and
biexciton de-excitonization energies $E_{\rm D}^{\rm X^\pm}$ and
$E_{\rm D}^{\rm XX}$ are
$E_{\rm D}^{\rm X^-} = E^{\rm X} - E^{\rm X^-}$, $E_{\rm D}^{\rm X^+}
= E^{\rm X} + E^{\rm h^+} - E^{\rm X^+}$, and $E_{\rm D}^{\rm XX} =
2E^{\rm X} - E^{\rm XX}$, where $E^i$ is the ground-state total energy for
complex $i$. These are the energies at which trion and biexciton peaks are
expected to appear relative to the exciton peak in the photoluminescence
spectrum of a quantum ring \cite{Young_2014}. The sign is such that for a free
trion or biexciton $E_{\rm D}>0$.  The binding energies---the energy difference
between a complex and its most energetically favorable daughter products,
bearing in mind that the holes are confined to the ring---are
$E_{\rm b}^{\rm X} = E^{\rm X} - E^{\rm h^+}$, $E_{\rm b}^{\rm X^-} =
E^{\rm X^-} - E^{\rm X}$, $E_{\rm b}^{\rm X^+} = E^{\rm X^+} - E^{\rm
  2h^+}$, and $E_{\rm b}^{\rm XX} = E^{\rm XX} - E^{\rm X^+}$, where
$E^{\rm 2h^+}$ is the energy of two holes confined to the same ring. The
binding energy determines the temperature at which a complex becomes
unstable against dissociation into smaller complexes.



\section{Computational Methodology}

The \textsc{casino} code \cite{Needs_2010} was used to perform DMC
calculations of the ground-state energies of excitons, trions, and
biexcitons in quantum-ring heterostructures. DMC is a stochastic
projection method that finds the ground-state component of a trial
wave function. In this work the trial wave function was optimized
using the variational Monte Carlo (VMC) method, in which many-body
expectation values are evaluated using Monte Carlo integration. The
trial wave function $\Psi_{\rm T}$ was of Slater-Jastrow form; e.g.,
for the biexciton:
\begin{equation}
\Psi_{\rm T}(\mathbf{R})=\exp[J({\bf R})]\phi_{\rm e}(\mathbf{r}_{\rm
  e_1})\phi_{\rm e}(\mathbf{r}_{\rm e_2}) \phi_{\rm h}(\mathbf{r}_{\rm
  h_1})\phi_{\rm h}(\mathbf{r}_{\rm h_2}),
\label{eq:trialwf}
\end{equation} 
where $\mathbf{R}=(\mathbf{r}_{\rm e_1},\mathbf{r}_{\rm
  e_2},\mathbf{r}_{\rm h_1},\mathbf{r}_{\rm h_2})$.
The hole orbital $\phi_{\rm h}$ was taken to be the exact ground-state
solution to the Schr{\"o}dinger equation for a single hole confined to
the ring:
\begin{equation} \phi_{\rm h}(\mathbf{r}) = \left[\frac{-J_0(\beta r)Y_0(\beta
r_{\rm i})}{J_0(\beta r_{\rm i})} + Y_0 (\beta r)\right]\cos\left(\frac{\pi
z}{2R_z}\right), \label{eq:hole_orb} \end{equation}
where $J_0$ and $Y_0$ are Bessel functions of the first and second
kind, respectively.  The constant $\beta$ is determined by imposing
the boundary condition $\phi_{\rm h}(r_{\rm o})=0$ numerically for
each ring size using the Newton-Raphson method; the other boundary
conditions are already satisfied by Eq.\ (\ref{eq:hole_orb}).
The electronic behavior is dominated by Coulomb attraction to the positively
charged ring together with hard-wall repulsion from the boundary of the ring.
At short range the electron orbital $\phi_{\rm e}$ linearly decreases to zero
on the ring boundary, while at long range the electron orbital decays
exponentially to keep the electrons localized to the ring; i.e., the behavior
is hydrogenic at long range. The electron orbital $\phi_{\rm e}$ enforces the
correct long- and short-range behavior, with the mid-range behavior 
determined by the Jastrow factor and variational freedom in the electron
orbital. The electron orbital was formed piecewise in eight regions about the
ring, with the functions in each region being matched at the boundaries to
ensure the value and gradient were smooth everywhere and the orbital was zero
inside the ring. See the Supplemental Material for the full form of the orbital.
The Jastrow exponent $J(\mathbf{R})$ included a pairwise sum of terms
of the form $u_{ij}(r) = \pm\mu_{ij} r/(1+c_{ij}r)$ for each pair of
particles $i$ and $j$ separated by distance $r$ with reduced mass
$\mu_{ij}$. The $+$ sign was used for particles with the same charge
and $-$ for particles with opposite charge. $c_{ij}$ is a variational
parameter, which was different for each particle-pair type.  This form
ensured the Kato cusp conditions were satisfied \cite{Kato_1957}. Other
one-, two-, and three-body polynomial terms were also included in the
Jastrow exponent; these were smoothly truncated at finite
range \cite{Drummond_2004,Lopez_2012}.
VMC energy minimization was used to optimize the trial wave
functions \cite{Umrigar_2007}. The fixed-node DMC algorithm is exact
for the ground-state energy of an exciton, trion, or biexciton,
because of the distinguishability of the particles, which leads to a
nodeless wave function.  Pairs of DMC calculations were performed with
time steps in a $1 : 4$ ratio and target configuration populations in
a $4 : 1$ ratio and the results were extrapolated linearly to zero
time step and infinite population. Charge densities were obtained by
binning the radial and axial coordinates of each of the particles
sampled during VMC and DMC calculations, cylindrically averaging, and
then calculating the extrapolated estimate. The errors in the VMC and
DMC estimates of the charge density ($\rho_{\rm VMC}$ and $\rho_{\rm
  DMC}$) are linear in the error in the trial wave function; however,
the error in the extrapolated estimate $2\rho_{\rm DMC} - \rho_{\rm
  VMC}$ is quadratic in the error in the trial wave
function \cite{Foulkes_2001}.



\section{Results and Discussion}

All energies and charge densities are reported for a ring composed of
GaSb surrounded by GaAs. The electron and hole masses are taken to be
$m_{\rm e}^*=0.063\ m_{\rm e}$ and $m_{\rm h}^*=0.4\ m_{\rm e}$,
respectively, where $m_{\rm e}$ is the bare electron mass.  The former
is the effective mass of an electron in bulk GaAs and the latter is
the mass of a heavy hole in bulk GaSb \cite{Levinshtein_1996}. The
permittivity is taken to be $\epsilon=12.9\ \epsilon_0$, where
$\epsilon_0$ is the permittivity of free space. This is the
permittivity of bulk GaAs \cite{Levinshtein_1996}. Data from
Ref.\ \onlinecite{Young_2012} were used to obtain experimentally
relevant values for the ring size; these values were $R_z=2.5$ nm
$=0.199\ a_0^*$, $r_{\rm i}=6$ nm $=0.479\ a_0^*$, and $r_{\rm o}=10$
nm $=0.799\ a_0^*$.  This geometry was used as the starting point for
our calculations; the aspect ratio $2R_z/(r_{\rm o}-r_{\rm i})$ of the
ring was then varied subject to the constraints that the volume of the
ring $2\pi R_z(r_{\rm o}^2-r_{\rm i}^2)$ was constant and the center
of the ring in the radial direction $(r_{\rm i}+r_{\rm o})/2$ was
fixed. A ring with aspect ratio much less than $1$ is akin to a thin
disc with a hole in the center, while a ring with aspect ratio much
greater than $1$ resembles a pipe.

\begin{figure}[!htpb]
\includegraphics[width=0.95\columnwidth]{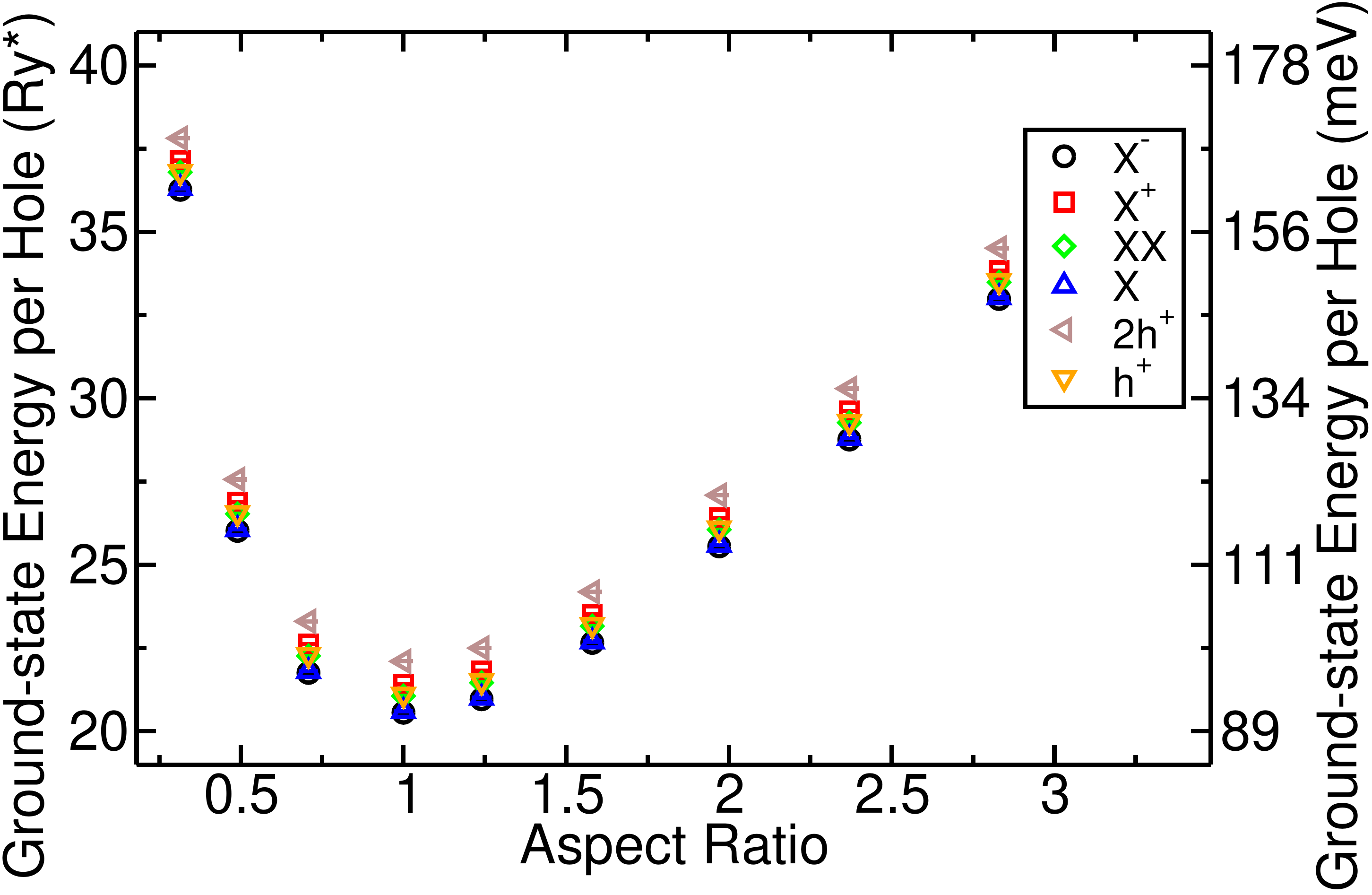}

\caption{Ground-state total energies per hole of a single hole (h$^+$), two
holes ($2{\rm h}^+$), an exciton (${\rm X}$), a negative trion (${\rm X}^-$), a
positive trion (${\rm X}^+$), and a biexciton (${\rm XX}$) in a quantum ring
plotted against the aspect ratio $2R_z/(r_{\rm o}-r_{\rm i})$ of the ring's
cross section. The mean radius and ring volume are appropriate for the
GaSb/GaAs quantum rings reported in Ref.\ \cite{Young_2012}. Error bars
are smaller than the size of the symbols. The exciton Rydberg $R_{\rm y}^*$ is
$4.45$ meV for the experimentally relevant geometry.}

\label{fig:gs_energies}
\end{figure}

The analytically evaluated variation in the hole energy against aspect
ratio is shown in Fig.\ \ref{fig:gs_energies}. The minimum energy
occurs when the cross-section is square; away from the minimum, the
energy goes roughly as $1/L^2$, where $L=\min\{2R_z,r_{\rm o}-r_{\rm
i}\}$.  Also shown in Fig.\ \ref{fig:gs_energies} are DMC ground-state
total energies per hole for $2{\rm h}^+$, ${\rm X}$, ${\rm X}^-$,
${\rm X}^+$, and ${\rm XX}$, all of which are bound.  These confirm
that the ground-state energies of the single-hole complexes (${\rm X}$
and ${\rm X}^-$) are very close to the energy of a single hole, while
the ground-state energies of the two-hole complexes (${\rm X}^+$ and
${\rm XX}$) are comparable with the energy of two confined holes. The
ground-state energies for single- and two-hole complexes vary slightly
differently as a function of aspect ratio due to the interaction
between the holes. The capacitive charging energy $E_{\rm CC}=E^{\rm
2h^+} - 2E^{\rm h^+}$ for the experimentally relevant ring geometry
\cite{Young_2012} is $E_{\rm CC}=8.8546(8)$ meV; this compares to an
experimentally measured value \cite{Hodgson_2013} of $E_{\rm
CC}=24(2)$ meV\@. STM images of quantum rings
\cite{Young_2012,Young_2014} suggest that the GaSb/GaAs interface is
not clean in practice. This disorder could lead to trapping of holes,
strongly affecting capacitive charging energies while having
relatively little effect on binding energies.

\begin{figure}[!htpb]
\includegraphics[width=0.95\columnwidth]{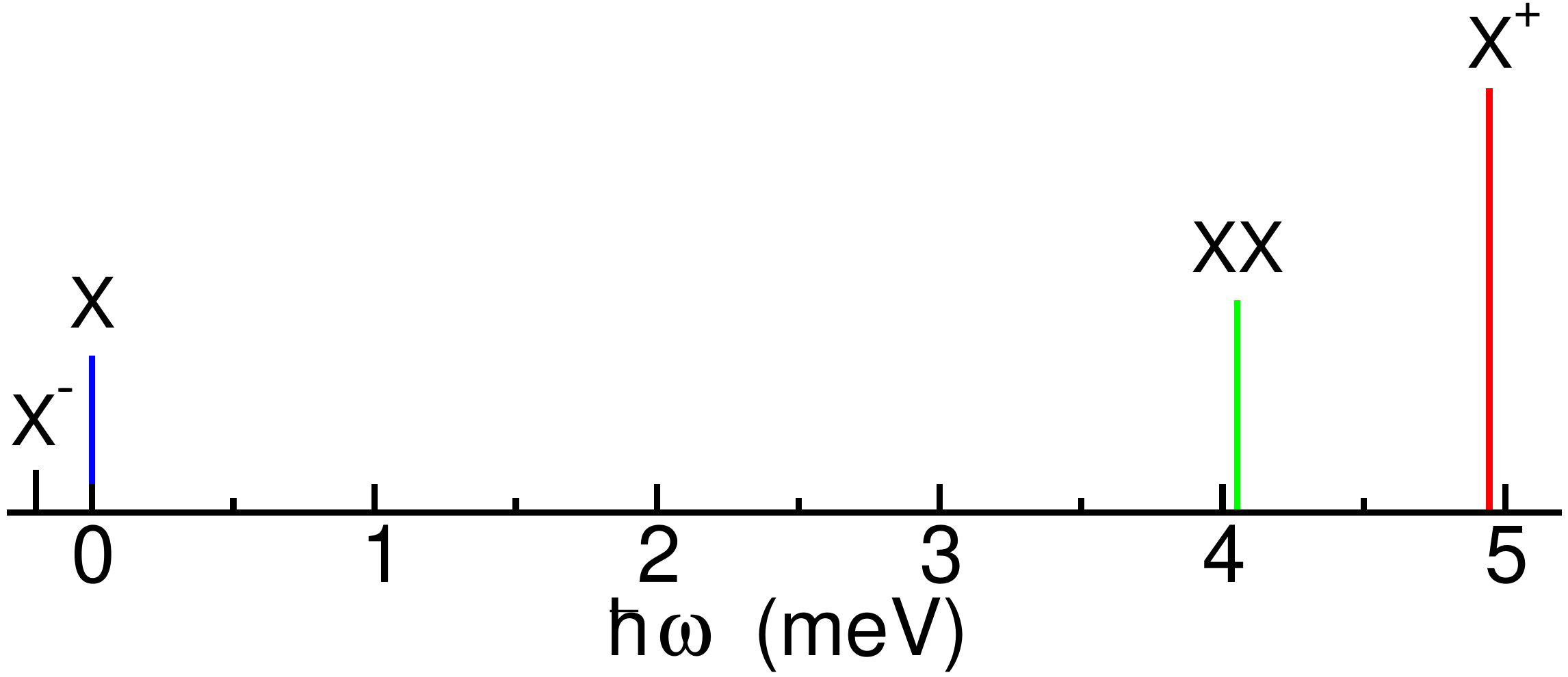}
\caption{Expected peak positions for the excitonic complexes in a
  photoluminescence spectrum relative to the exciton peak, for a model
  of the quantum rings reported in Ref.\ \cite{Young_2012}. The
  peak heights represent the relative stability of the complexes.}
\label{fig:pl_spec} 
\end{figure}

\begin{figure}[!htpb]
\includegraphics[width=0.95\columnwidth]{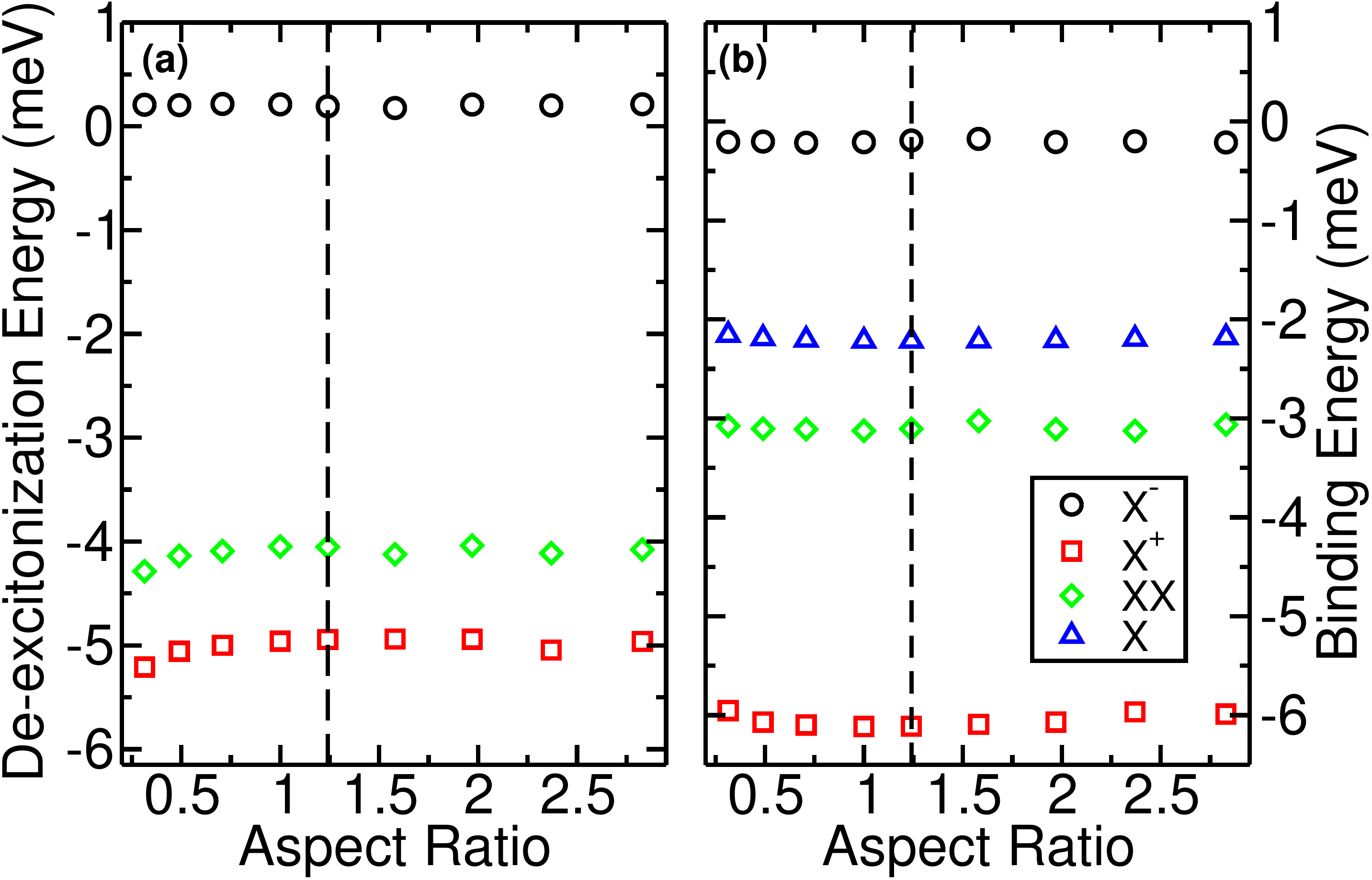}
\caption{(a) De-excitonization energies and (b) binding energies 
  against the aspect ratio $2R_z/(r_{\rm o}-r_{\rm i})$ of a quantum 
  ring's cross-section for different charge-carrier complexes. 
  The mean radius and ring volume are appropriate for the GaSb/GaAs 
  quantum rings reported in Ref.\ \cite{Young_2012}. Error bars 
  are smaller than the size of the symbols. The dashed lines shows the
  experimentally relevant aspect ratio \cite{Young_2012}.}
\label{fig:deenergy}
\end{figure}

The de-excitonization energies for the trions and biexciton in the
geometry modeling the quantum rings described in Ref.\
\cite{Young_2012} can be found in Table \ref{tab:energies}. The
de-excitonization energy is positive for ${\rm X}^-$, but negative for
${\rm X}^+$ and ${\rm XX}$. The negative de-excitonization energy is a
result of the large energy penalty when two holes are confined to the
same ring; e.g., two excitons on two separate quantum rings would be
the energetically preferred four-particle state rather than a
biexciton on a single ring. The expected positions of these peaks in a
photoluminescence spectrum are shown in Fig.\ \ref{fig:pl_spec}. The
${\rm X}^-$ peak is very close to the ${\rm X}$ peak, while the peaks
for ${\rm X}^+$ and ${\rm XX}$ are separated from the ${\rm X}$ peak
by a few meV\@. The heights of the peaks indicate the relative
stability of the complexes, using binding energy data from Table
\ref{tab:energies}. Experimental work has not yet progressed to the
point where excitonic complex peak positions have been identified. The
only published work showing sharp lines in the photoluminescence
spectra of GaSb/GaAs quantum rings is Ref.\ \cite{Young_2014}; however
the spectra in this work would likely contain peaks from many, highly
positively charged rings, making a direct comparison with theoretical
values difficult.  The de-excitonization energy is plotted against the
aspect ratio of the cross-section of the ring for ${\rm X}^-$, ${\rm
X}^+$, and ${\rm XX}$ in Fig.\ \ref{fig:deenergy}(a). For each complex
it can be seen that there is some slight change in the
de-excitonization energy as a function of aspect ratio.  The
de-excitonization energies are largely independent of the aspect
ratio, and hence exact shape of the ring, somewhat justifying the use
of a ring with a rectangular cross-section in our model.  Furthermore,
the energetic effects of the slight interpenetration of the electron
and hole orbitals are likely to be well described by a slight
renormalization of the cross-section of the ring; however the effects
of such small changes in the cross-section appear to be small.

\begin{table}
\begin{center}
\caption{De-excitonization $E_{\rm D}$ and binding $E_{\rm b}$ energies for
excitonic complexes in the quantum-ring geometry modeling the samples described
in Ref.\ \cite{Young_2012}. \label{tab:energies}}
\begin{tabular}{lr@{.}lr@{.}lr@{.}lr@{.}l}
\hline \hline

Complex & \multicolumn{2}{c}{$E_{\rm D}/R_{\rm y}^*$} &
\multicolumn{2}{c}{$E_{\rm b}/R_{\rm y}^*$} & \multicolumn{2}{c}{$E_{\rm D}$
(meV)} & \multicolumn{2}{c}{$E_{\rm b}$ (meV)}\\ \hline

${\rm X}$ & \multicolumn{2}{c}{0} & $-0$&$5004(6)$  & \multicolumn{2}{c}{0} &
$-2$&$226(3)$\\

${\rm X}^-$ & $+0$&$0446(4)$ & $-0$&$0446(4)$ & $+0$&$199(2)$ & $-0$&$199(2)$\\

${\rm X}^+$ & $-1$&$111(2)$ & $-1$&$379(2)$ & $-4$&$944(7)$ & $-6$&$137(7)$\\

${\rm XX}$ & $-0$&$911(2)$ & $-0$&$701(2)$ & $-4$&$052(8)$ & $-3$&$11(1)$ \\

\hline \hline
\end{tabular}
\end{center}
\end{table}

The binding energies for each complex are shown in Table \ref{tab:energies} for
the experimentally relevant geometry \cite{Young_2012}. The ${\rm X}$ binding
energy is about half the value for a free ${\rm X}$ due to the exclusion of the
electron from the ring.  As expected, ${\rm X}^-$ is the most weakly bound
(against dissociation into a free electron and a neutral exciton), while ${\rm
X}^+$ is the most stable (against removal of an electron from a ring of charge
of $+2e$). From these binding energies the temperatures up to which the
complexes are stable are $26$, $2.3$, $71$, and $36$ K for ${\rm X}$, ${\rm
X}^-$, ${\rm X}^+$, and ${\rm XX}$, respectively. As with the de-excitonization
energies, the binding energy depends weakly on the aspect ratio of the ring's
cross-section, but again these differences are much smaller than the
differences in binding energy between complexes: see Fig.\
\ref{fig:deenergy}(b). Therefore, the binding energy appears to be largely
independent of the exact shape of the cross-section of the ring for a given
ring volume and mean radius.

\begin{figure}[!b]
\centering
\includegraphics[width=0.95\columnwidth]{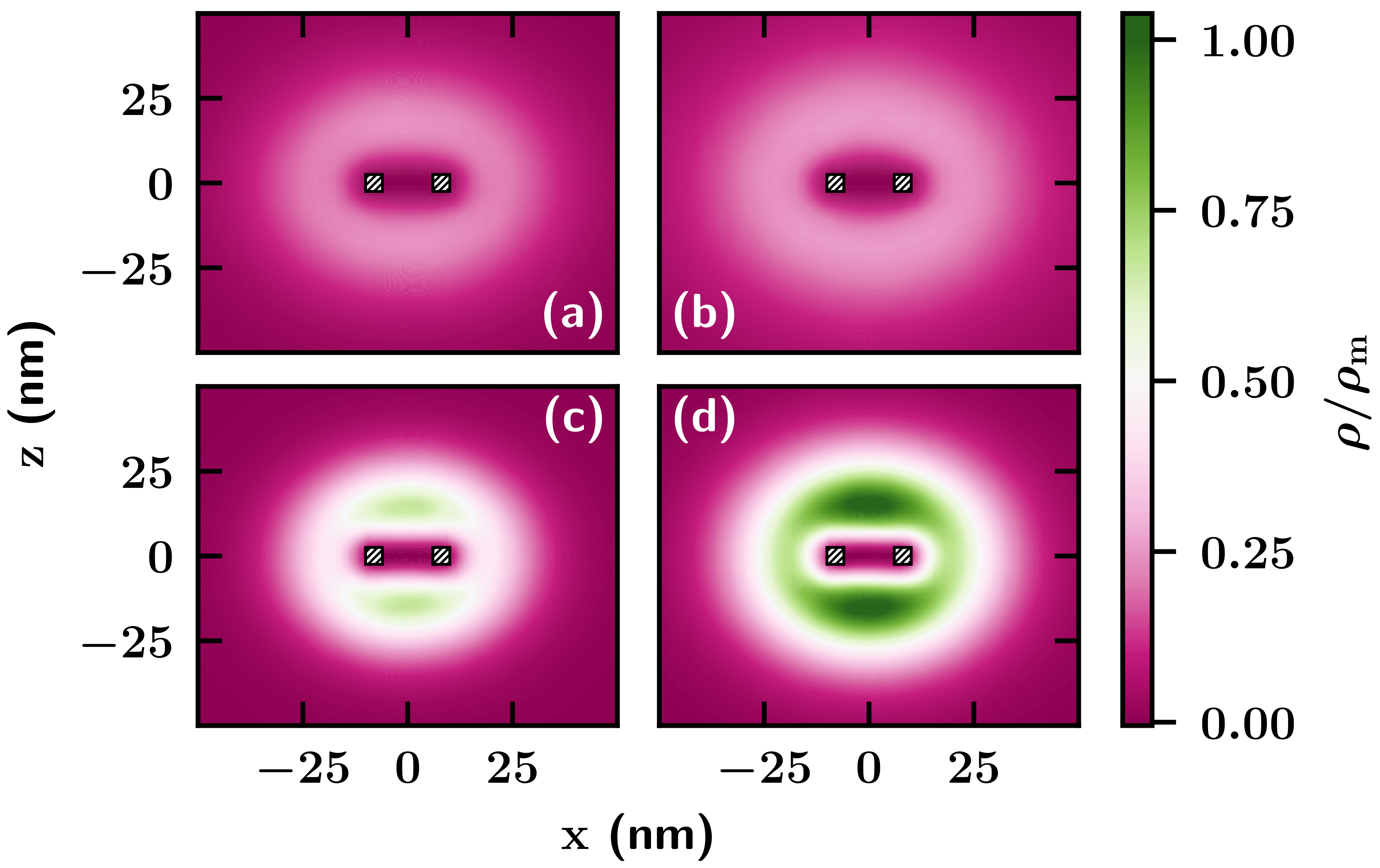}
\caption{Electronic charge density $\rho$ for (a) an exciton, (b) a
negative trion, (c) a positive trion, and (d) a biexciton in the experimentally
relevant quantum-ring geometry \cite{Young_2012}. The shaded regions represent
the ring and $\rho_m$ is the maximum density across all four plots.
 The free exciton Bohr radius is $a_0^*=12.5$ nm.}
\label{fig:chargedensity}
\end{figure}

Plots of the electronic charge density for each complex in the experimentally
relevant geometry are shown in Fig.\ \ref{fig:chargedensity}.  The electrons
form a diffuse halo around the ring, with negligible charge density in the
ring's central cavity. The kinetic-energy cost of localizing in the ring's
cavity significantly exceeds the gain in electrostatic potential energy.
Correlation effects further reduce the probability of finding multiple
electrons inside the ring's cavity. ${\rm XX}$ and ${\rm X}^+$ are
the most localized complexes, as reflected in their relatively large binding
energies shown in Table \ref{tab:energies}. These two-hole complexes have
slightly higher electronic charge densities in the regions directly above and
below the center of the ring compared to the regions to the left and right of
the ring.  STM images of the electronic density of states in Ref.\
\cite{Young_2012} suggest the electrons are localized to the ring's
cavity, which does not agree with the results presented here. However, in
the STM experiments the sample is cleaved in the $x$-$z$ plane. This is a
drastic modification to the system, which is not described by our model.
It is plausible that the reduced screening
and hence smaller free exciton Bohr radius in the cleaved
system allows electrons to localize within, rather than above
or below, the quantum ring.
 
The sensitivity of the ${\rm XX}$ binding energy to various parameters
is presented in Table \ref{tab:sensitivity}. The ${\rm XX}$ binding
energy depends most strongly on the electron effective mass, and is
relatively insensitive to the hole effective mass, relative
permittivity, ring volume, and mean ring radius.  Our conclusions are
robust against reasonable uncertainties in model parameters.

\begin{table}[!t]
\begin{center}
\caption{Sensitivity of the biexciton binding energy to the electron
  and hole effective masses $m_{\rm e}$ and $m_{\rm h}$, the relative
  permittivity $\epsilon$, the ring volume $V$, and the mean radius of
  the ring $r_{\rm m}$.\label{tab:sensitivity}}
\begin{tabular}{r@{.}lr@{.}lr@{.}lr@{.}lr@{.}l}
\hline \hline \\ [-1.0ex]

\multicolumn{2}{c}{$\partial E_{\rm b}^{\rm XX}/\partial m_{\rm e}$} &
\multicolumn{2}{c}{$\partial E_{\rm b}^{\rm XX}/\partial m_{\rm h}$} &
\multicolumn{2}{c}{$\partial E_{\rm b}^{\rm XX}/\partial \epsilon$} &
\multicolumn{2}{c}{$\partial E_{\rm b}^{\rm XX}/\partial V$} &
\multicolumn{2}{c}{$\partial E_{\rm b}^{\rm XX}/\partial r_{\rm m}$}  \\
\multicolumn{2}{c}{(meV/$m_{\rm e}$)} & 
\multicolumn{2}{c}{(meV/$m_{\rm e}$)} &
\multicolumn{2}{c}{(meV)} &
\multicolumn{2}{c}{(meV/nm$^3$)} & 
\multicolumn{2}{c}{(meV/nm)} \\ \hline
$-7$&$4(3)$ & $-0$&$20(4)$ & $0$&$39(1)$ & $0$&$0004(2)$ & $0$&$07(2)$ \\
\hline \hline
\end{tabular}
\end{center}
\end{table}

Kehili \textit{et al.} \cite{Kehili_2018}\ have recently investigated
excitons in GaSb rings in GaAs quantum wells using the effective-mass
approximation, modeling the ring with a finite potential, and
including strain effects due to lattice-constant mismatch. In their
work the interaction between charge carriers is described by a Hartree
mean-field approximation, in contrast to the complete treatment of
correlation effects used here. Nevertheless, their electronic charge
density is qualitatively consistent with our results. Their ${\rm X}$
binding is slightly larger than our value reported in Table
\ref{tab:sensitivity}, however, partly due to their use of slightly
different effective masses and mean ring radii. A DMC calculation of
the ${\rm X}$ binding energy using the same ring geometry and
effective masses as Kehili \textit{et al.}\ gives $E_{\rm
  b}^{\rm X}=2.695(2)$ meV, which is comparable with the binding
energy of about 2.6 meV that they report for a GaAs well of width of
$40$ nm (the largest well width they consider). The ${\rm X}$ binding
energies reported by Kehili \textit{et al.}\ do not appear to have
converged with respect to well width at this point, however, and it looks
as if they will be significantly smaller than the DMC exciton binding
energy in the limit of large well width. This is consistent with the
fact that, by the variational principle, Hartree theory underestimates
the magnitude of the ${\rm X}$ binding energy.



\section{Conclusion}

In conclusion, total energies of excitonic complexes in type-II
quantum-ring heterostructures are dominated by the confinement energy
of the holes in each complex. The de-excitonization energy is positive
for ${\rm X}^-$ as would be the case for a free trion; however,
for ${\rm X}^+$ and ${\rm XX}$ this energy is negative due to the
energy penalty associated with confining two holes in the same
ring. ${\rm X}^-$ is the least stable of the complexes studied;
it is predicted to be stable only at temperatures below $2.3$ K,
while the most stable complex, ${\rm X}^+$, endures up to
$71$ K\@.  De-excitonization and binding energies were shown to be
largely independent of the aspect ratio at fixed ring volume and mean
radius, suggesting these energies may also be fairly independent of
the precise shape of the cross-section of the ring.  The electrons
form a halo around the outside of the ring, with a low density in the
central cavity. This reflects the fact that the ring size is
comparable with the free exciton Bohr radius. Furthermore, 
${\rm X}^+$ and ${\rm XX}$ are the most tightly bound complexes,
with a preference for the electrons to position themselves above and
below the ring. For ${\rm X}^-$, the electronic charge density
is much more diffuse, consistent with its very small binding energy.


\begin{acknowledgments}
D.M.T.\ and R.J.H.\ are fully funded by the Graphene NOWNANO CDT (EPSRC Grant
No.\ EP/L01548X/1). Computer resources were provided by Lancaster University's
High End Computing Cluster. We acknowledge useful discussions with B.\
Garikipati.
\end{acknowledgments}


\bibliography{qr}

\end{document}